\documentclass[conference]{IEEEtran}
\IEEEoverridecommandlockouts
\usepackage{cite}
\usepackage{amsmath}
\usepackage{amssymb,amsfonts}
\usepackage{amsthm}
\usepackage{array}
\usepackage{algorithm}
\usepackage[noend]{algpseudocode}
\usepackage{amsthm}
\usepackage[colorlinks=true,pdfpagemode=none,urlcolor=blue,linkcolor=blue,citecolor=violet,pdfstartview=FitH]{hyperref}
\usepackage{graphicx}
\usepackage{multirow}
\usepackage{subcaption}
\usepackage{paralist}
\usepackage{enumitem}
\usepackage{textcomp}
\usepackage{xcolor,colortbl}
\usepackage{fancyhdr}

\newtheorem{theorem}{Theorem}[section]
\newtheorem{claim}[theorem]{Claim}

\newtheorem{definition}[theorem]{Definition}
\newtheorem{lemma}{Lemma}[theorem]

\newcommand{\THISWORK}{{\fontfamily{lmss}\selectfont 
TEACUPS}}
\newcommand{\pluseq}{\mathrel{+}=}

\newcommand{\tempout}[3]{d^+_#1[#2,#3]}
\newcommand{\tempin}[3]{d^-_#1[#2,#3]}

\newcommand{\outlist}[3]{\Lambda^+_#1[#2, #3]}
\newcommand{\inlist}[3]{\Lambda^-_#1[#2, #3]}

\def\BibTeX{{\rm B\kern-.05em{\sc i\kern-.025em b}\kern-.08em
    T\kern-.1667em\lower.7ex\hbox{E}\kern-.125emX}}
\begin{document}

\fancypagestyle{firstpage}
{
    \fancyhead[C]{Authors' Version; to appear in the Proceeding of International Conference on Data Mining (ICDM 2024).}
}
\renewcommand{\headrulewidth}{0pt}

\newcolumntype{P}[1]{>{\centering\arraybackslash}p{#1}}
\newcolumntype{C}[1]{>{\centering\arraybackslash}m{#1}}   
\newcolumntype{R}[1]{>{\raggedleft\arraybackslash}m{#1}}  
\newcolumntype{L}[1]{>{\raggedright\arraybackslash}m{#1}}  

\colorlet{Green1}{green!25}
\colorlet{Blue}{blue!25}
\definecolor{Green2}{HTML}{99d8c9}


\newcommand{\EX}{\mathbf{E}}
\newcommand{\prob}{\mathbf{Pr}}
\newcommand{\eqdef}{:=}

\newcommand{\eps}{\varepsilon}


\newcommand{\cA}{{\cal A}}
\newcommand{\cB}{\mathcal{B}}
\newcommand{\cC}{{\cal C}}
\newcommand{\cD}{\mathcal{D}}
\newcommand{\cE}{{\cal E}}
\newcommand{\cF}{\mathcal{F}}
\newcommand{\cG}{\mathcal{G}}
\newcommand{\cH}{{\cal H}}
\newcommand{\cI}{{\cal I}}
\newcommand{\cJ}{{\cal J}}
\newcommand{\cL}{{\cal L}}
\newcommand{\cM}{{\cal M}}
\newcommand{\cP}{\mathcal{P}}
\newcommand{\cQ}{\mathcal{Q}}
\newcommand{\cR}{{\cal R}}
\newcommand{\cS}{\mathcal{S}}
\newcommand{\cT}{{\cal T}}
\newcommand{\cU}{{\cal U}}
\newcommand{\cV}{{\cal V}}
\newcommand{\cX}{{\cal X}}

\newcommand{\Sec}[1]{\hyperref[sec:#1]{Section~\ref*{sec:#1}}} 
\newcommand{\Eqn}[1]{\hyperref[eq:#1]{(\ref*{eq:#1})}} 
\newcommand{\Fig}[1]{\hyperref[fig:#1]{Figure\,\ref*{fig:#1}}} 
\newcommand{\Tab}[1]{\hyperref[tab:#1]{Table\,\ref*{tab:#1}}} 
\newcommand{\Thm}[1]{\hyperref[thm:#1]{Theorem\,\ref*{thm:#1}}} 
\newcommand{\Fact}[1]{\hyperref[fact:#1]{Fact\,\ref*{fact:#1}}} 
\newcommand{\Lem}[1]{\hyperref[lem:#1]{Lemma\,\ref*{lem:#1}}} 
\newcommand{\Prop}[1]{\hyperref[prop:#1]{Prop.~\ref*{prop:#1}}} 
\newcommand{\Cor}[1]{\hyperref[cor:#1]{Corollary~\ref*{cor:#1}}} 
\newcommand{\Conj}[1]{\hyperref[conj:#1]{Conjecture~\ref*{conj:#1}}} 
\newcommand{\Def}[1]{\hyperref[def:#1]{Definition~\ref*{def:#1}}} 
\newcommand{\Alg}[1]{\hyperref[algo:#1]{Algorithm~\ref*{algo:#1}}} 
\newcommand{\Ex}[1]{\hyperref[ex:#1]{Example.~\ref*{ex:#1}}} 
\newcommand{\Clm}[1]{\hyperref[clm:#1]{Claim~\ref*{clm:#1}}} 
\newcommand{\Step}[1]{\hyperref[step:#1]{Step~\ref*{step:#1}}} 

\newcommand{\Yunjie}[1]{\textcolor{blue}{Yunjie: #1}}
\newcommand{\Omkar}[1]{\textcolor{blue}{Omkar: #1}}

\title{Accurate and Fast Estimation of Temporal Motifs using Path Sampling}
\author{\IEEEauthorblockN{Yunjie Pan}
\IEEEauthorblockA{
\textit{University of Michigan}\\
Ann Arbor, USA \\
panyj@umich.edu}
\and
\IEEEauthorblockN{Omkar Bhalerao}
\IEEEauthorblockA{
\textit{University of California, Santa Cruz}\\
Santa Cruz, USA \\
obhalera@ucsc.edu}
\and
\IEEEauthorblockN{C. Seshadhri}
\IEEEauthorblockA{
\textit{University of California, Santa Cruz}\\
Santa Cruz, USA \\
sesh@ucsc.edu}
\and
\IEEEauthorblockN{Nishil Talati}
\IEEEauthorblockA{
\textit{University of Michigan}\\
Ann Arbor, USA \\
talatin@umich.edu}
\and
}

\maketitle
\thispagestyle{firstpage}
\pagestyle{plain}
\begin{abstract}
Counting the number of small subgraphs, called motifs, is a fundamental problem in social network analysis and graph mining. Many real-world networks are directed and temporal, where edges have timestamps. Motif counting in directed, temporal graphs is especially challenging because there are a plethora of different kinds of patterns. Temporal motif counts reveal much richer information and there is a need for scalable algorithms for motif counting.

A major challenge in counting is that there can be trillions of temporal motif matches even with a graph with only millions of vertices. Both the motifs and the input graphs can have multiple edges between two vertices, leading to a combinatorial explosion problem. Counting temporal motifs involving just four vertices is not feasible with current state-of-the-art algorithms.

We design an algorithm, \THISWORK, that addresses this problem using a novel technique of temporal path sampling. We combine a path sampling method with carefully designed temporal data structures, to propose an efficient approximate algorithm for temporal motif counting. \THISWORK\ is an unbiased estimator with provable concentration behavior, which can be used to bound the estimation error. 
For a Bitcoin graph with hundreds of millions of edges, \THISWORK\ runs in less than 1 minute, while the exact counting algorithm takes more than a day. We empirically demonstrate the accuracy of \THISWORK\ on large datasets, showing an average of 30$\times$ speedup (up to 2000$\times$ speedup) compared to existing GPU-based exact counting methods while preserving high count estimation accuracy.
\end{abstract}
\section{Introduction}


Mining small subgraph patterns, referred to as \textit{motifs}, is a central problem in network analysis~\cite{milo2002network}.
Motif mining plays a critical role in understanding the structure and function of complex systems encoded as graphs~\cite{shen2002network, alon2007network, newman2003structure}. 
There is a rich body of work on mining motifs in static graphs~\cite{iyer2018asap, jha2015path, pinar2017escape, wang2017moss, jain2020power, bressan2018motif} (refer to the tutorial~\cite{seshadhri2019scalable} for details).
Most real-world phenomena representing networks, \textit{e.g.,} social interaction, communication, are dynamic, where edges are created with timestamps.
Static graphs are built by omitting crucial temporal information.

Temporal edges between nodes are tuples $(u, v, t)$, where $u$ and $v$ are source and destination nodes, and $t$ is the timestamp of the edge. 
Edges with timestamps capture richer information compared to static edges~\cite{kovanen2011temporal, pan2011path}. A graph with such temporal edges is called a \emph{temporal graph}, and patterns in such graphs are called \emph{temporal motifs}.
Temporal motifs are useful in user behavior characterization on social/communication networks~\cite{kovanen2011temporal, kovanen2013temporal, lahiri2007structure, paranjape2017motifs}, detecting fraud in financial transaction networks~\cite{hajdu2020temporal,liu2023temporal}, and characterizing the structure and function of biological networks~\cite{pan2011path,hulovatyy2015exploring}.
Furthermore, local motif counts are useful to resolve symmetries and improve the expressive power of GNNs~\cite{bouritsas2022improving, liu2023temporal,rossi2020temporal}.

\begin{figure}
    \centering
    \includegraphics[width=0.85\columnwidth]{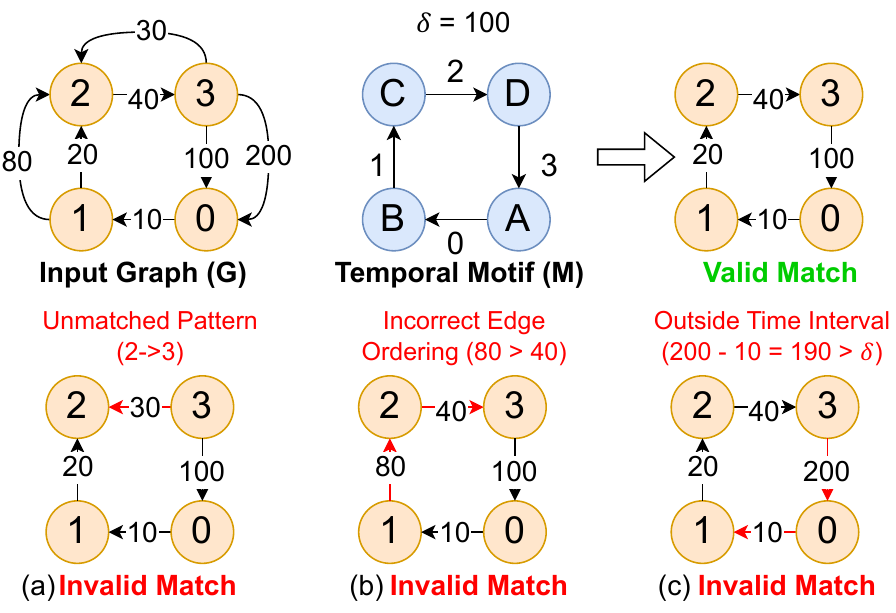}
    \vspace{-3mm}
    \caption{An example of mining a temporal motif (M) from a input graph (G)}
    \label{fig:motif-match-example}
\end{figure}

\subsection{Formal Problem Definition} \label{subsection:problem_definition}

The input is a directed temporal graph $G = (V(G), E(G))$, with $n$ vertices and $m$ edges. Each temporal edge is a tuple $e=(u,v,t)$ where $u$ and $v$ are vertices in the temporal graph, and $t$ is a positive integer timestamp. Note that the graph is directed, and there can be many edges between the same $u$ and $v$. For temporal edge $e$, we use $t(e)$ to denote its timestamp.
For convenience, we think of time in seconds.

We formally define a temporal motif and a match, following Paranjape \textit{et al.}~\cite{paranjape2017motifs}. (Our match definition is non-induced, since we only match edges. This is consistent with past work.)

\begin{definition} \label{def:temp-motif} A \emph{temporal motif} is a triple $M=(H,\pi, \delta)$ where (i) $H = (V(H), E(H))$ is a directed pattern multi-graph, $(ii)$ $\pi$ is a permutation on the edges of $H$, and (iii) $\delta$ is a positive integer.

The permutation $\pi$ specifies the time ordering of edges, and $\delta$ specifies the length of time interval for all edges.
\end{definition}

\begin{definition} \label{def:temp-match} Consider an input temporal graph $G = (V(G),E(G))$ and a temporal pattern $M = (H,\pi,\delta)$. An \emph{$M$-match} is a 1-1 map $\phi:V(H) \to V(G)$ satisfying the following conditions.
\begin{asparaitem}
    \item (Matching the pattern) The map $\phi$ matches the edges of $H$. Formally, $\forall (u,v) \in E(H)$, $(\phi(u),\phi(v)) \in E(G)$. For convenience, for edge $e \in E(H)$, we use $\phi(e)$ to denote the match in the pattern.
    \item (Edges ordered correctly) The timestamps of the edges in the match follow the ordering $\pi$. Formally, $\forall e, e' \in E(H)$, $\pi(e) < \pi(e')$ iff $t(\phi(e)) < t(\phi(e'))$.
    \item (Edges in time interval) All edges of the match occur within $\delta$ time. $\forall e, e' \in E(H)$, $|t(\phi(e)) - t(\phi(e'))| \leq \delta$.
\end{asparaitem}
\end{definition}

\begin{table}[t]
    \scriptsize
    \caption{Summary of notation}
    \label{tab:notation}
    \begin{tabular}{L{2.3cm}|L{5.6cm}}
         Symbol & Definition \\ \hline\hline
         $\delta$ & maximum time window\\
         $M=(H,\pi, \delta)$ & a directed temporal motif  with $\delta$ time window \\
         $G=(V(G), H(G))$ & a directed temporal graph \\
         $S$ & the wedge or 3-path chosen from $M$\\
         $P$ & the sampled wedge or 3-path from $G$ that maps to $S$\\
         $w_{e, \delta}$ & the sampling weight of edge $e$ in $G$\\
         $W_{\delta}$ & the total number of $\delta$-centered wedge or 3-path\\
    \end{tabular}
    
\end{table}

For convenience, we summarize the symbols and definitions in \Tab{notation}. Some symbols are defined later.

As a walkthrough example in Figure~\ref{fig:motif-match-example}, we want to mine a temporal motif $M$ with $\delta=100$ from an input graph $G$. We show one valid match and three invalid matches. Those matches are invalid due to: (a) the edge direction between nodes 2 and 3 is opposite to the pattern in M; (b) edge (1, 2, 80) occurs after edge (2, 3, 40), which violates the edge ordering constraints in M; (c) the edge (3, 0, 200) happens 190 seconds after the first edge (0, 1, 10), exceeding the maximum time window $\delta$.



\begin{figure}
    \captionsetup[subfigure]{justification=centering}
    \centering
    \begin{subfigure}{0.23\linewidth}
       \includegraphics[width=0.88\linewidth]{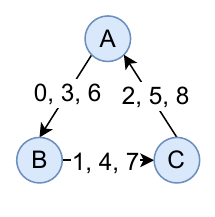} 
       \caption{$M_\text{3-0}$}
       \label{fig:M3-0}
    \end{subfigure}
    \begin{subfigure}{0.23\linewidth}
    \includegraphics[width=0.88\linewidth]{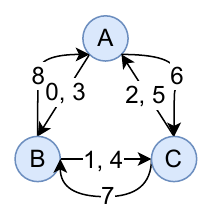} 
       \caption{$M_\text{3-1}$}
       \label{fig:M3-1}
    \end{subfigure}
    \begin{subfigure}{0.23\linewidth}
    \includegraphics[width=0.88\linewidth]{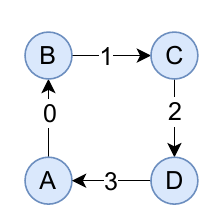} 
       \caption{$M_\text{4-0}$}
       \label{fig:M4-0}
    \end{subfigure}
    \begin{subfigure}{0.23\linewidth}
       \includegraphics[width=0.88\linewidth]{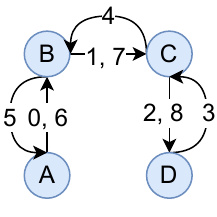} 
       \caption{$M_\text{4-1}$}
       \label{fig:M4-1}
    \end{subfigure}

    \begin{subfigure}{0.23\linewidth}
       \includegraphics[width=0.88\linewidth]{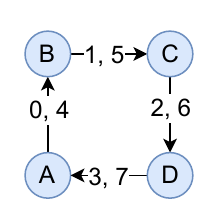}  
       \caption{$M_\text{4-2}$}
       \label{fig:M4-2}
    \end{subfigure}
    \begin{subfigure}{0.23\linewidth}
       \includegraphics[width=0.88\linewidth]{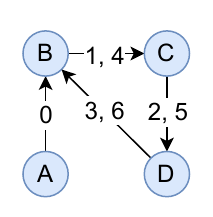} 
       \caption{$M_\text{4-3}$}
       \label{fig:M4-3}
    \end{subfigure}
    \begin{subfigure}{0.23\linewidth}
        \includegraphics[width=0.88\linewidth]{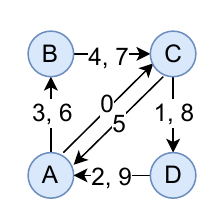} 
       \caption{$M_\text{4-4}$}
       \label{fig:M4-4}
    \end{subfigure}
    \begin{subfigure}{0.23\linewidth}
       \includegraphics[width=0.88\linewidth]{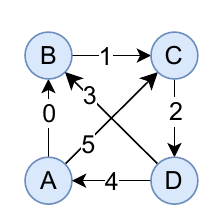} 
       \caption{$M_\text{4-5}$}
       \label{fig:M7}
    \end{subfigure}
    \caption{A subset of connected temporal motifs.}
    \label{fig:temp_motifs}
\end{figure}

In Figure~\ref{fig:temp_motifs}, we show 3- and 4-vertex temporal motifs.
In our definition, we also allow multiple edges between two vertices. For example, pattern $M_{4-4}$ involves 10 edges. Every pair of vertices has 2 edges. A match needs to find 10 edges satisfying the edge pattern, directions, and timestamp orderings. 

\begin{figure}[bt]
    \centering
    \begin{subfigure}{\linewidth}
        \includegraphics[width=\linewidth]{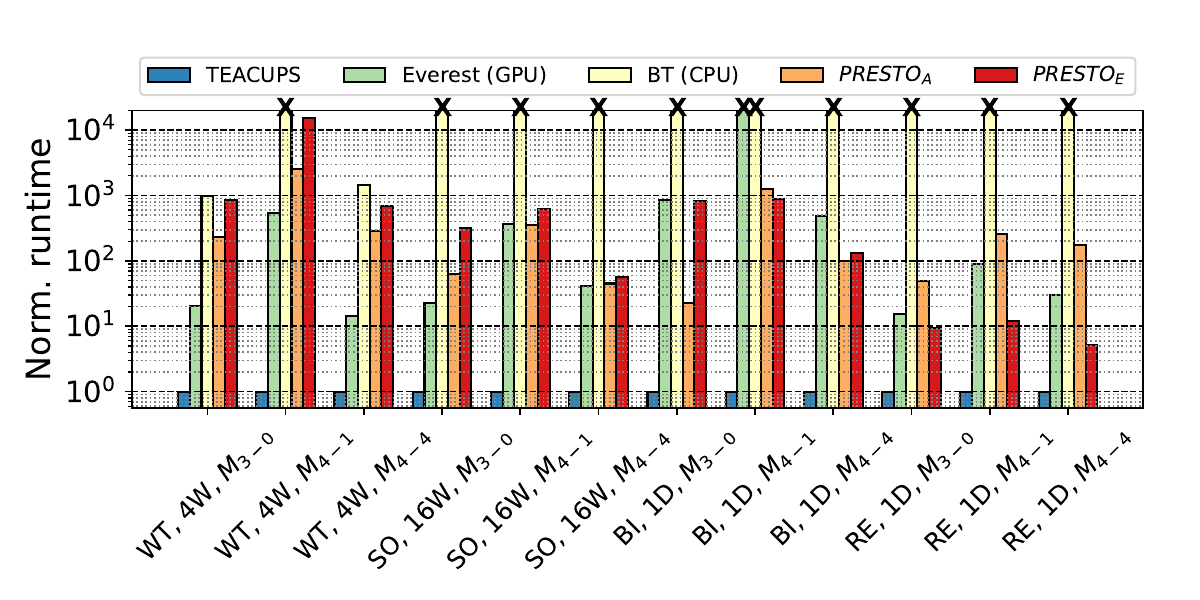}
        \caption{Runtime of prior works normalized to \THISWORK(ours). \THISWORK\ has up to 1e4$\times$ speedup compared to previous works.}
        \label{fig:runtime}
    \end{subfigure}
    \begin{subfigure}{\linewidth}
        \centering
        \includegraphics[width=\linewidth]{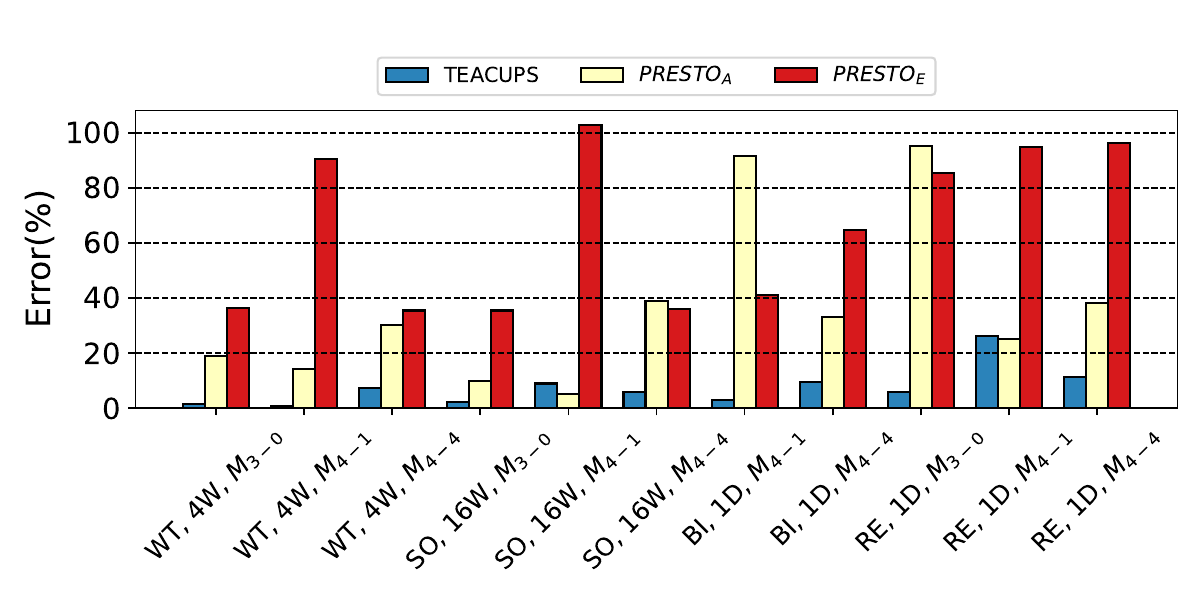}
        \caption{Relative error (\%) of approximate algorithms. \THISWORK(ours) is almost always the most accurate while always the fastest. The errors for (BI, 1D, $M_\text{3-1}$) are missing because the exact count is unavailable (Everest GPU cannot finish in 1 day). }
        \label{fig:error}
    \end{subfigure}
    \caption{Normalized runtime of exact algorithms (Everest GPU and BT CPU) and approximate algorithms (\THISWORK, $PRESTO_A$, and $PRESTO_B$) on various datasets, time ranges and motifs. Except Everst (GPU) which runs on an NVIDIA A40 GPU, all other experiments run on CPU with 32 threads.}
\end{figure}


\subsection{Challenges in Mining Temporal Motifs} \label{sec:challenges}

There are \textbf{two} major challenges where temporal motif counting distinguishes itself from static versions. 
\textit{First}, a temporal motif involves ordering relations between the timestamps of edges. For example, Figure~\ref{fig:motif-match-example}(b) shows an invalid match because of violating the edge ordering constraints. 
Most of the best static motif counting techniques exploit motif structure and static graph properties like degeneracy~\cite{seshadhri2014wedge,pinar2017escape,bressan2018motif,seshadhri2019scalable}. These specific techniques do not work when imposing ordering constraints on edges. Currently, the specialized temporal motif counting are only practical for simple temporal triangles (3-vertex motifs)~\cite{paranjape2017motifs,kumar20182scent,pashanasangi2021faster}. 
\textit{No current method (general and specialized) can get 4-vertex motif count results for the Bitcoin graph (100M temporal edges) even within \emph{one day} on commodity GPU platform.}

\textit{Second} challenge is the \textit{combinatorial explosion problem} due to multi-graphs (both the input graph and motif). In input graphs, there can be thousands of temporal edges between the same pair of edges. (A static directed graph only has two possible edges.) The edge multiplicity of the input graph itself already causes a large amount of search space and the number of matched instances. For Bitcoin graph with 110 million edges, there are trillion copies of a simple temporal 4-cycle ($M_\text{4-0}$). 
However, the introduction of multi-graph motifs further compounds this explosion, making the search space for temporal multi-graph motifs orders of magnitude larger than their simpler counterparts. 
The number of instances of temporal multi-graph 4-cycle ($M_\text{4-2}$) in the Bitcoin input graph reaches hundreds of trillion, two orders of magnitude larger than the simple 4-cycle. 
Methods based on enumeration or exploration, no matter how well designed, cannot avoid this massive computation~\cite{paranjape2017motifs,mackey2018chronological}.

\textbf{To summarize}, there are \emph{numerous} temporal patterns (like Figure~\ref{fig:temp_motifs}) involving just 3- and 4-vertices, so we cannot hope to define specialized algorithms for motifs. 
At the same time, general methods are often based on neighborhood exploration that suffer a massive computational explosion because of temporal edge multiplicity.

This motivates the central question behind this work: \emph{How can we design scalable algorithms for temporal motif counting that can go beyond triangles and extend to patterns with multiple parallel edges between any pair of vertices?}


To address these challenges, this work presents the first scalable algorithm for counting temporal patterns involving \textit{up to four vertices and any number of edges} in large networks.
While the best prior algorithms only work for patterns with up to three vertices~\cite{seshadhri2014wedge,pinar2017escape,bressan2018motif,seshadhri2019scalable}, we take a major step forward to support complex pattern matching use-cases.
An extension of our algorithm to count motifs with more than four vertices involves generalizing path sampling techniques to spanning trees, which is out of the scope of this paper, and is left for future work.

\subsection{Main Contributions} \label{sec:contributions}

In this paper, we design the \emph{Temporal Explorations Accurately Counted Using Path Sampling} algorithm, called \THISWORK. 
\THISWORK\ is a fast and accurate open-source\footnote{Code available at \url{https://github.com/pyjhzwh/TEACUPS_ICDM}.} temporal path sampling algorithm that estimates temporal motif counts.
Below, we summarize our contributions.

\noindent{\bf The Concept of Sampling Temporal Paths.} Our main conceptual contribution is the introduction of temporal path sampling. Many of the best static motif counting methods use path or tree sampling~\cite{seshadhri2014wedge,jha2015path,wang2017moss,turk2019revisiting}. 
We show how to generalize the method of wedge or $3$-path sampling for temporal graphs. 
There are significant challenges since direction and temporal orderings create many different kinds of wedges/$3$-paths for a single static wedge/$3$-path.
We couple the randomized sampling methods with carefully designed temporal data structures. This leads to an all-purpose temporal sampler, that can sample paths with any ordering and time interval constraints. Based on this sampler, we design the \THISWORK\ algorithm. \emph{All} temporal motifs on $4$ vertices contain a temporal $3$-path, and \emph{all} 3-vertex motifs contain a temporal wedge. \THISWORK\ is a randomized algorithm that can be applied to all such motifs to estimate motif counts.

In the static case~\cite{seshadhri2014wedge}, there is an easy way to extend the sampled 3-path into an instance of the motif by checking if the edge exists. However, this does not work for temporal motif counting due to multi-graph. We present an efficient algorithm, which determines the number of instances of the target motif induced by the sampled 3-path in time, linear in the multiplicity of the edges. 

\noindent\textbf{Counting Temporal Multi-Graph Motifs.} In extending motif patterns to encompass multi-graphs, we uncover more intricate transactional or communicational features in real-world scenarios. Although input graphs often take the form of multi-graphs, prior research has largely neglected motifs within this context. Existing general algorithms~\cite{paranjape2017motifs, mackey2018chronological} show diminished efficiency in this domain, while algorithms designed for rapid processing ~\cite{seshadhri2014wedge, kumar20182scent, pashanasangi2021faster} are ineffectual for temporal multi-graph motif counting. 
To the best of our knowledge, this is the first work that specifically focuses on motif patterns that are multi-graphs.

\noindent\textbf{Provable Correctness with Bounded Error.}
\THISWORK\ always gives unbiased estimates. 
Additionally, we rigorously bound the approximation error using techniques from randomized algorithms. We apply
concentration inequalities to prove that \THISWORK\ accurately estimates
to temporal motif counts.

\noindent\textbf{Scalable Runtime.} We perform an empirical analysis of \THISWORK\ on numerous real-world temporal datasets. \THISWORK\ is extremely fast. 
Figure~\ref{fig:runtime} shows the highlight of runtime for the state-of-the-art algorithms (exact and approximate). 
We give results for numerous datasets and time windows. Across \emph{all} experiments, \THISWORK\ is $10$-$1000\times$ faster than existing works. 
For example, \THISWORK\ takes \textit{less than 1 minute} on a Bitcoin graph with 110M edges, where the motif count is over a trillion. The exact count algorithm takes more than one day, and approximate methods either have high error rates or are at least $10\times$ slower.
Even compared to the exact count algorithm that runs on an NVIDIA GPU, our algorithm that runs on a CPU  exhibits an average of 30$\times$ speedup.
Additionally, \THISWORK\ exhibits a near-linear runtime scalability with CPU threads.




\noindent\textbf{Empirical Accuracy.}
We empirically validate \THISWORK\ on a variety of datasets and conduct convergence experiments.
\THISWORK\ consistently achieves a relative error of 
$<10\%$ for the majority of motifs. In all cases, \THISWORK\ has a lower or comparable error to existing sampling methods, while being at least 10$\times$ faster, as shown in Figure~\ref{fig:error}.


\section{Related Work} \label{sec:related}
\noindent\textbf{Static Motif Mining.}
There are various works on mining static motifs, including the exact count or enumeration methods~\cite{chen2022efficient, jamshidi2020peregrine, jamshidi2021deeper,mawhirter2019graphzero, mawhirter2019automine, shi2020graphpi, teixeira2015arabesque, wei2022stmatch,li2024fast}, and approximate techniques~\cite{schank05finding, seshadhri2014wedge, jha2015path, liu2019sampling, wang2020efficient, tsourakakis2009doulion, ahmed2014graph, turk2019revisiting, pavan2013counting, wang2017moss, jain2017fast, ye2022lightning, ye2023efficient, yang2021efficient, bressan2019motivo}. Although static motif mining can serve as a first step for temporal motif mining~\cite{paranjape2017motifs}, it is shown that it causes orders of magnitude redundant work, which calls for efficient algorithms tailored to the temporal motif mining problem.

\noindent\textbf{Exact Temporal Motif Counting.}
The problem of temporal motif mining is first formally described by Paranjape \textit{et al.}~\cite{paranjape2017motifs}, where they propose an algorithm to enumerate and count the temporal motif instances.
A more recent exact counting algorithm~\cite{mackey2018chronological} proposed a backtracking algorithm by enumerating all instances on chronologically sorted edges. 
Everest~\cite{yuan2023everest} supports the backtracking exact counting algorithm on GPUs with system-level optimizations that improve performance by an order of magnitude.
There are other general temporal motif count algorithms~\cite{sun2019tm,min2023time}. A critical challenge with the exact counting algorithms is scalability with respect to graph and motif sizes as enumeration all possible matches takes a large portion of memory and compute resources.
A few algorithms focus on mining specific motifs~\cite{boekhout2019efficiently,kumar20182scent,pashanasangi2021faster,gao2022scalable}. These solutions, however, are limited to simpler motifs and do not address multi-graph motifs.

\noindent\textbf{Approximate Temporal Motif Counting.}
To address the scalability issue, Interval-Sampling~\cite{liu2019sampling} proposed a sampling algorithm by partitioning times into intervals. PRESTO~\cite{sarpe2021presto} improved the sampling method by leveraging uniform sampling based on Liu \textit{et al.}~\cite{liu2019sampling}. 
Edge-Sampling~\cite{wang2020efficient} estimates the total number of temporal motifs by exactly counting the local motifs from uniformly sampled temporal edges. 
These approaches, however, have two drawbacks. 
First, they use uniform sampling while the input graphs are skewed, affecting the accuracy of estimates.
Second, their dependency on exact backtracking count algorithm~\cite{mackey2018chronological} on processing sampled edge subsets necessitates examining every instance, reducing their efficiency. 
Additionally, an online, single-pass sampling algorithm was developed by Ahmed \textit{et al.}~\cite{ahmed2021online}. Also, Oden~\cite{sarpe2021oden} a sampling-based work that could count multiple motifs that has the same underlying static structure.

\section{Temporal 3-path Sampling}

We first define the specific class of temporal $3$-paths that we sample from. Recall that edges of the input graph $G = (V, E)$ are tuples of the form $(u,v,t)$ where $u$ and $v$ are vertices, and $t$ is a timestamp (in seconds). They are sorted by timestamp.

We introduce several crucial definitions, focusing on the concepts of $\delta$-centered wedges and $\delta$-centered 3-paths. In this work, our focus is on using $\delta$-centered 3-paths to obtain accurate estimates of motif-counts for patterns on 4 vertices. This is primarily because counting motifs on 4-vertices is more challenging than triangles, and all of the ideas presented for this setting can be easily generalized to patterns involving 3 vertices.
\begin{definition} \label{def:wedge} A \emph{$\delta$-centered wedge} is a pair of edges $e_1, e_2$ with the following properties. Let $e_1 = (u,v,t(e_1))$. Then, $e_2$ is incident to $v$. Furthermore, $|t(e_1) - t(e_2)| \leq \delta$. We call $e_1$ the "base" edge.
\end{definition}

Next, we introduce the notion of a $\delta$-centered $3$-path. 
\begin{definition} \label{def:3path} A \emph{$\delta$-centered $3$-path} is a sequence
of three edges $e_1, e_2, e_3$ with the following properties. Let $e_2 = (u,v,t(e_2))$. Then, $e_1$ is incident to $u$ and $e_3$ is incident to $v$. Furthermore, $|t(e_1) - t(e_2)| \leq \delta$ and $|t(e_3) - t(e_2)| \leq \delta$.
\end{definition}
In particular, a $\delta$-centered $3$-path is a sequence of three "connected" edges with both end edges within time $\delta$ of the center edge. Our key observation is that such temporal $3$-paths can be sampled rapidly.

A few things to note. There is no $\delta$ constraint between $t(e_1)$ and $t(e_3)$. Also, we allow $e_1$ and $e_3$ to potentially intersect (at a vertex), and even allow $e_1$ and $e_3$ to contain both $u$ and $v$. Technically, this forms a homomorphism of a $3$-path. We choose the definition above for easy sampling while maintaining some temporal constraints.

We introduce a further categorization of temporal $3$-paths, into 16 classes indexed by 4-bit binary tuples. Specifically, our classes are denoted $\langle \alpha_1, \alpha_3, \beta_1, \beta_3 \rangle$, where $\alpha_1, \alpha_3 \in \{+,-\}$ and $\beta_1, \beta_3 \in \{<,>\}$. $\alpha_i, \beta_i$ represent a constraint on the edge $e_i$ of a $\delta$-centered $3$-path.

Instead of giving a general definition for all classes, we define the $\langle -,+,<,> \rangle$ class. All other classes are defined analogously.

\begin{definition} \label{def:3path-class} The class of $\langle -,+,<,> \rangle$ $\delta$-centered $3$-paths contains all paths of the following form. Let $(e_1, e_2, e_3)$ be a $\delta$-centered $3$-path and $e_2 = (u,v,t)$ be the center edge. Then $e_1$ is an inedge ($-$) of $u$ and $e_3$ is an outedge ($+$) of $v$. 
Also, $t(e_1) < t(e_2)$ and $t(e_3) > t(e_2)$.
\end{definition}

We stress that there is no condition between $e_1$ and $e_3$. Note that we can define $\langle +,+,<,< \rangle$ classes and so forth. 

We introduce a key definition of multiplicity that are required to express runtime bounds.

\begin{definition} \label{def:multiplicity} Given two vertices $u, v$ and timestamps $t < t'$, the \emph{multiplicity} $\sigma_{u,v}[t,t']$ is the number of edges $(u,v,t'')$ such that $t'' \in [t,t']$.
We denote the maximum $\delta$-multiplicity as $\sigma_\delta$, 
defined as $\max_{u,v,t}$ $\sigma_{u,v}[t,t+\delta]$. 
\end{definition}

The principle idea behind \THISWORK\ is to accurately estimate the number of connected 3- or 4-vertex temporal motifs by sampling a subset of temporal wedges or 3-paths, whose edges must satisfy certain temporal constraints, and then determine the motif counts. The algorithm comprises three primary phases: 1) preprocessing to get sampling weights, 2) sampling wedges/3-paths, and 3) deriving desired motif counts from the sampled wedges/3-paths. As mentioned before, in this work, we will focus on 3-path sampling and how it can be used to derive motif counts for connected 4-vertex patterns. 

We describe the procedures in the following sections. For convenience, the graph $G = (V,E)$ is assumed to be a global variable accessible to all procedures.

\subsection{Preprocessing and Sampling Procedure}
\label{sec:path-sampling-algorithm}

Our algorithm to estimate the number of connected 4-vertex temporal motifs is primarily based on 3-path sampling. Observe that every 4-vertex motif in Figure \ref{fig:temp_motifs} contains at least one 3-path. We choose one 3-path as the anchor $S$.
Our algorithm and analysis work for all classes of 3-paths. For the sake of readability, we will present the path sampling algorithm for the $\langle -,+,<,> \rangle$ class. All other classes can be sampled analogously.

Our main module is a uniform random sampler for $3$-paths from a specific class. Let's first set some definitions.

\begin{definition} [Temporal outlists and degrees, $\outlist{v}{t}{t'}$, $\tempout{v}{t}{t'}$]
\label{def:in_out_edges}
Given a vertex $v$ and natural numbers $t,\delta$, the \emph{temporal outlist} $\outlist{v}{t}{t'}$ is defined as the set of outedges $(v,w,t'')$, where $w$ is an arbitrary vertex
and $t'' \in [t,t']$. We similarly define the temporal inlist as $\inlist{v}{t}{t'}$.

The \emph{temporal out-degree} $\tempout{v}{t}{t'}$ denotes the size of $\outlist{v}{t}{t'}$. The \emph{temporal in-degree} $\tempin{v}{t}{t'}$ is the size of $\inlist{v}{t}{t'}$.

\begin{definition} [$3$-path counts $w_{e,\delta}$ and $W_\delta$] \label{def:3path-count}
Fix $\delta$. For a temporal edge $e = (u,v,t)$, $w_{e,\delta}$ denotes the number of $\delta$-centered $\langle -,+,<,> \rangle$ 3-paths, where $e$ is the center edge. 

We use $W_\delta := \sum_e w_{e,\delta}$ to be the total number of $\delta$-centered $\langle -,+,<,> \rangle$ 3-paths.
\end{definition}

\end{definition}

The following claim is central to the algorithm and analysis.

\begin{claim} \label{clm:count} 
$w_{e,\delta} = \tempin{u}{t-\delta}{t} \cdot \tempout{v}{t}{t+\delta}$.
\end{claim}

\begin{proof} Consider any pair of edges $(x,u,t_x)$ and $(v,w,t_w)$ where
$x,w$ are arbitrary vertices, $t_x \in [t-\delta,t]$ and $t_w \in [t,t+\delta]$. Then, these edges form a $\langle +,-,<,> \rangle$ $\delta$-centered $3$-path with $e$ at the center. The number of edges $(x,u,t_x)$ where $t_x \in [t-\delta,t]$ is precisely the indegree $\tempin{u}{t-\delta}{t}$. Similarly, the number of edges $(v,w,t_w)$ where $t_w \in [t,t+\delta]$ is the outdegree $\tempout{v}{t}{t+\delta}$. The product of these degrees gives the total number of desired $3$-paths, which is $\tempin{u}{t-\delta}{t} \cdot \tempout{v}{t}{t+\delta}$.
\end{proof}

Note that we can count $\langle \alpha,\alpha',<,> \rangle$ 3-paths with the formula
$d^\alpha_u[t-\delta,t] \cdot d^{\alpha'}_v[t,t+\delta]$. (To change the $<,>$ indicators
in the class, we would focus on different time interval.)

We now describe the actual procedure that samples temporal $3$-paths. We assume that the graph $G$ is stored as an adjacency list of out-edges \emph{and} in-edges, where each list is sorted by timestamp. So we can perform binary search by time among the edges incident to a vertex. After we select a 3-path $S$ from the motif $M$, we do the following two procedures.

\begin{algorithm}[t]
\caption{: \textsc{Preprocess}($\delta$)}
\label{algo:preprocess}
\begin{flushleft}
\textbf{Input}: Time $\delta$ (code assumes $\langle -,+,<,> \rangle$ class)\\
\textbf{Output}: Sampling probability $p_{e,\delta}$ for each edge $e\in E$, and the total sampling weight $W_\delta$.
\begin{algorithmic}[1]
\For{$v \in V$}
\State In the in-neighbor list of $v$, binary search for the times $t-\delta$ and $t$. Store the difference in indices to be $\tempin{v}{t-\delta}{t}$.
\State In the out-neighbor list of $v$, binary search for the times $t$ and $t+\delta$. Store the difference in indices to be $\tempout{v}{t}{t+\delta}$.
\EndFor
\State $W_\delta = 0$
\For{$e = (u,v,t) \in E$}
\State $w_{e, \delta} = \tempin{u}{t-\delta}{t} \cdot \tempout{v}{t}{t+\delta}$
\State $W_\delta \pluseq w_{e,\delta}$
\EndFor
\For{$e \in E$}
\State $p_{e,\delta} = w_{e,\delta} / W_\delta$
\EndFor
\end{algorithmic}
\end{flushleft}
\end{algorithm}

First is the \textsc{Preprocess} procedure (\Alg{preprocess}) that constructs a distribution based on $w_{e,\delta}$ values. It crucially uses \Clm{count}. The set of values $\{p_{e,\delta}\}$ forms a distribution over the edges. The next procedure, \textsc{Sample} (in \Alg{sample}), computes a random $3$-path based on the distribution constructed in the preprocessing step.

\begin{algorithm}[t]
\caption{: \textsc{Sample}$(\{p_{e,\delta}\},\delta)$}
\label{algo:sample}
\begin{flushleft}
\textbf{Input}: Sampling distribution $\{p_{e,\delta}\}$ from \textsc{Preprocess}, and time  $\delta$ \\
\textbf{Output}: Uniform random $\langle -,+,<,> \rangle$ $\delta$-centered $3$-path $P$
\end{flushleft}
\begin{algorithmic}[1]
\State Pick edge $e = (u, v, t)$ with probability $p_{e, \delta}$. 
\State In the in-neighbor list of $u$, binary search for the times $t-\delta$ and $t$ to get $\inlist{u}{t-\delta}{t}$.
\State Pick uniform random edge $e_1$ from $\inlist{u}{t-\delta}{t}$.
\State In the out-neighbor list of $v$, binary search for the times $t$ and $t+\delta$ to get $\outlist{v}{t}{t+\delta}$.
\State Pick uniform random edge $e_3$ from $\outlist{v}{t}{t+\delta}$\\
\Return $P=(e_1, e, e_3)$ 
\end{algorithmic}
\end{algorithm}

\begin{claim} \label{clm:runtime-sample} The procedure \textsc{Preprocess} runs in time $O(m\log m)$. The procedure \textsc{Sample} runs in time $O(\log m)$ per sample.
\end{claim}

\begin{proof} We first give the running time of \textsc{Preprocess}. Observe that it performs four binary searches for each edge (two each in the out-neighbors and in-neighbors). The total running time is $O(m\log m)$. $m$ is the number of edges in $G$.

For \textsc{Sample}, the first step is to sample from the distribution given by $\{p_{e,\delta}\}$ values. This can be done using a binary search, which takes $O(\log m)$ time. After that, it performs two binary searches and two random number generations. So the total running time is $O(\log m)$.
\end{proof}

Finally, we show that \textsc{Sample} is a bonafide uniform random sampler.

\begin{lemma} \label{lem:sample} The procedure \textsc{Sample} generates a uniform random $\langle -,+,<,> \rangle$ $\delta$-centered $3$-path.
\end{lemma}

\begin{proof} Consider a $\langle -,+,<,> \rangle$ $\delta$-centered $3$-path $(e_1, e_2, e_3)$. Let $e_2 = (u,v,t)$. Then $e_1$ is an in-edge of $u$ with timestamp in $[t-\delta,t]$. Also, $e_3$ is an out-edge of $v$ with timestamp in $[t,t+\delta]$.

The probability of sampling $e_2$ is $w_{e_2,\delta}/W_\delta$. Conditioned on this sample,
the probability of sampling $e_1$ is precisely $1/|\inlist{u}{t-\delta}{t}|$, which
is $1/\tempin{u}{t-\delta}{t}$. Similarly, the probability of sampling $e_3$ is $1/\tempout{v}{t}{t+\delta}$. Multiplying all of these, we get the probability of sampling the entire $3$-path $(e_1, e_2, e_3)$.
Applying \Clm{count}, the probability is 
\begin{eqnarray*} 
& & \frac{w_{e_2,\delta}}{W_\delta} \cdot \frac{1}{\tempin{u}{t-\delta}{t}} \cdot \frac{1}{\tempout{v}{t}{t+\delta}} \\
& = & \frac{\tempin{u}{t-\delta}{t} \cdot \tempout{v}{t}{t+\delta}}{W_\delta} \cdot \frac{1}{\tempin{u}{t-\delta}{t} \cdot \tempout{v}{t}{t+\delta}} = \frac{1}{W_\delta}
\end{eqnarray*}
Hence, the probability of sampling $(e_1, e_2, e_3)$ is $1/W_\delta$, which corresponds to the uniform distribution. (By definition, the total number of $\langle -,+,<,> \rangle$ $\delta$-centered $3$-paths is $W_\delta$.)
\end{proof}

\subsection{Counting Motifs That Extend a 3-path}
\label{sec:check_motif}

The previous section discussed how to sample a 3-path. Once we have a sampled 3-path $P$, we need to determine if it can be ``extended" to count a desired motif. We describe the algorithms formally as the procedures \textsc{CheckMotif} and \textsc{ListCount} in high-level. 

We first denote the list of edges $(u, v, t)$ in $G$ as $L_{u,v}$. (No constraints on $t$.) 
We aim to count the temporal motif $M = (H,\pi,\delta)$, using $M_\text{4-1}$ from \Fig{temp_motifs} as an example. To do this, we first select a specific 3-path within the motif, referred to as the \emph{anchor} $S$. This anchor includes the motif edge with the earliest timestamp, labeled as time order 0. For $M_{4-1}$, we choose the 3-path $S$ consisting of the edges labeled $0, 1, 2$ (the sequence $(A,B,0), (B,C,1), (C,D,2)$).

\begin{algorithm}[t]
\caption{: {\textsc{CheckMotif}$(P,M)$}}
\label{algo:checkMotif}
\begin{flushleft}
\textbf{Input}: 3-path $P=\{e_1,e_2,e_3\}$, where $e_1=\{u',u,t_1\}$, $e_2=\{u,v,t_2\}$, $e_3=\{v, v',t_3\}$, temporal motif $M = (H,\pi,\delta)$ with a chosen anchor path\\
\textbf{Output}: Motif count $cnt$
\end{flushleft}
\begin{algorithmic}[1]
\If{($\{u', u, v, v'\}$ not all distinct) or ($t_3 > t_1 + \delta$)}
\State \Return $0$ 
\EndIf
\State For every edge $e$ of $M$ that is not on the anchor path, use binary search to find the sublist (in $G$) $E_f$ of potential matches. 
\State Number these lists as $L_1, L_2, \cdots L_l$ in the time ordering.
\State \Return \textsc{ListCount}$(L_1, L_2, \cdots L_l)$
\end{algorithmic}
\end{algorithm}

The \textsc{CheckMotif} procedure (in \Alg{checkMotif}) takes a sampled 3-path $P$ from $G$ as input and returns the count of temporal motif matches to $M$ where $P$ maps to anchor $S$. \textsc{CheckMotif} first checks if the sampled 3-path $P$ is a valid match to anchor $S$ in $M$. The checks are simple: the 3-path must span 4 vertices, and the edge timestamps should have the right order and be within $\delta$ timesteps. 

For motif $M$, There could be many different matches of $M$ that have $P$ mapped to the anchor path. For the example of $M_\text{4-1}$, we need to find matches to the edges $(A,B,6)$, $(B,A,5)$, $(B,C,7)$, $(C,B,4)$, $(D,C,3)$ and $(C,D,8)$. The adjacency lists are sorted by time. So, using binary search, we can find a sublist of edges in $G$ matching those edges whose timestamp is within $\delta$ of $P$ \emph{and} is larger than all timestamps in $P$. Let us denote these sublists as $L_{(A,B)}, L_{(B,A)}, L_{(B,C)}, L_{(C,B)}, L_{(D,C)}$ and $L_{(C,D)}$ respectively. 

We have a further constraint to satisfy. 
We need to count the number of tuples of edges $(e_{1}, e_{2}, e_{3}, e_{4}, e_{5}, e_{6})$ from those sublists that maintain this order. To efficiently count these matches without enumerating all possibilities, we employ a method combining pointer traversals with dynamic programming, as described in \Alg{listcount}.

\begin{algorithm}[t]
\caption{: \textsc{ListCount}($L_1, L_2, \dots, L_l$)}
\label{algo:listcount}
\begin{flushleft}
\textbf{Input}: $l$ sorted lists $L_1, L_2, \dots, L_l$\\
\textbf{Output}: The number of $l$ tuples $(l_1, l_2, l_3, \dots, l_l)$ for which the following holds : 1) $l_i \in L_i$ for all $1 \leq i \leq l$ and 2) $l_1 \leq l_2 \leq  \dots \leq l_k$
\begin{algorithmic}[1]
\State init $cnt_1 = \{1\} \times |L_1|$ \Comment{$cnt_r$ has the same length as $L_r$}
\For{$ r \in [2, l]$}
    \For{$i \in [0, |L_r|)$} \Comment{$i$ is the pointer to $L_r$}
        \State $j = 0$ \Comment{$j$ is the pointer to $L_{r-1}$}
        \State $cursum = 0$
        \While{$j < |L_{r-1}|$ and $L_{r-1}[j] \leq L_r[i]$}
        \State $j \pluseq 1$
        \State $cursum \pluseq cnt_{r-1}[j]$ 
        \EndWhile 
        \State $cnt_r[i] = cursum$
        \State delete $cnt_{r-1}$ 
    \EndFor
\EndFor
\State \Return $\sum cnt_{r}$
\end{algorithmic}
\end{flushleft}
\end{algorithm}


\begin{claim}\label{clm:listcount-runtime}
Given $l$ sorted lists $L_1, L_2, \dots, L_l$, \textsc{ListCount} runs in time $O(\sum_{r = 1}^{l}|L_r|)$, where $|L_r|$ is the length of $L_r$. 
\end{claim}

\subsection{The Overall Estimation Procedure} \label{sec:est}

\begin{algorithm}[t]
\caption{\textsc{Estimate}$(M,k)$}
\label{algo:overall}
\begin{flushleft}
\textbf{Input}: Temporal motif $M = (H,\pi,\delta)$, and sample number $k$ \\
\textbf{Output}: Motif count estimate $\widehat{C}$
\end{flushleft}
\begin{algorithmic}[1]
\State Choose an anchor 3-path of $M$ that has the smallest timeorder edge.
Let the 3-path be of type $\langle \alpha_1, \alpha_3, \beta_1, \beta_3 \rangle$.
\State $\{p_{e,\delta}\}, W_\delta = \textsc{Preprocess}(\delta)$ (for $\langle \alpha_1, \alpha_3, \beta_1, \beta_3 \rangle$ class paths)
\label{algo:overall-preprocess}
\State Initialize $cnt = 0$
\For{$i \in [1,k]$}
    \State $P_i = \textsc{Sample}(\{p_{e,\delta}\}, \delta)$  (for $\langle \alpha_1, \alpha_3, \beta_1, \beta_3 \rangle$ class) \label{algo:overall-sample}
    \State $cnt = cnt + \textsc{CheckMotif}(P_i, M)$ \label{algo:overall-checkMotif} 
\EndFor
\State \Return $\widehat{C} = (cnt/k)\cdot W_\delta$
\end{algorithmic}
\end{algorithm}

We now bring the pieces together with our main procedure \textsc{Estimate} (in \Alg{overall}).
There are some convenient notations to set up for the proof. We use the random variable $X_i$ to denote the output of \textsc{CheckMotif}$(P_i, M)$, which is the number of motif counts extended from the sampled path $P_i$. Observe that each path sample $P_i$ is chosen independently, so all the $X_i$'s are independent. We stress that while the random variables all depend on the graph, we assume that the graph is fixed. The only randomness is over the sampling of the path (in \textsc{Sample}), and the samples are independent of each other. Hence, the $X_i$'s are iid random variables.


\begin{claim} \label{clm:est} Let $C$ denote the number of $M$-matches in $G$.
Then $\EX[\widehat{C}] = C$.
\end{claim}

\begin{proof} Every $M$-match in $G$ has a unique anchor path. Let $\cP$ be the set of $\delta$-centered 3-paths, of the class determined by the anchor path. For a path $P \in \cP$, let $C_P$ be the number of $M$-matches where the path $P$ is the anchor. Since each match
contains a single anchor path, $\sum_{P \in \cP} C_P = C$. Observe that many $C_P$ values may be zero.

The output of \textsc{CheckMotif}$(P_i, M)$ is precisely the number of $M$-matches containing $P_i$ as an anchor; thus, $X_i = C_{P_i}$. Note that $cnt = \sum_{i \leq k} C_{P_i}$. By linearity of expectation, $\EX[cnt] = \sum_{i \leq k} \EX[C_{P_i}]$.

By the properties of \textsc{Sample} (\Lem{sample}), the path $P_i$ is a uniform random element of $\cP$. Recall (\Def{3path-count}) that $W_\delta$ is the total size of $\cP$. Hence, $\EX[C_{P_i}] = \sum_{P \in \cP} C_P/W_\delta = C/W_\delta$. So $\EX[cnt] = kC/W_\delta$.

The final output $\widehat{C} = (cnt/k)\cdot W_\delta$, so $\EX[\widehat{C}] = (W_\delta/k) \cdot \EX[cnt] = (W_\delta/k) \cdot kC/W_\delta = C$.
\end{proof}

We now show the concentration behavior of $\widehat{C}$. We will need the multiplicative Chernoff bound stated below.

\begin{theorem} \label{thm:chernoff} [Theorem 1.1 of~\cite{DuPa-book}] Let $Y_1, Y_2, \ldots, Y_k$ be independent random variables in $[0,1]$, and let $Y = \sum_{i \leq k} Y_i$. Let $\eps \in (0,1)$. Then, 
$$ \prob[|Y - \EX[Y]| > \eps \EX[Y]] \leq 2 \exp(-\eps^2 \EX[Y]/3)$$
\end{theorem}
Using the Chernoff bound, we prove the following theorem.
\begin{theorem} \label{thm:est} Suppose the temporal motif $M$ has $r$ edges that are not in selected wedge or 3-path $S$. ($r = |E(M)|-|V(M)|+1)$. Let the error probability be denoted $\gamma > 0$. Let the number of samples $k$ be at least $3\sigma^{r}_\delta W_\delta \ln(\gamma/2)/(C\eps^2)$. Then, $\prob[|\widehat{C} - C| > \eps C] < \gamma$.
\end{theorem}

\begin{proof} The random variables $X_i$ are potentially larger than $1$, so we can only apply \Thm{chernoff} after suitable scaling. For any path $P \in \cP$, let us bound the maximum value of $C_P$, which is the number of $M$-matches where $P$ is the anchor. Observe that the matches are determined the edges \emph{other than} $P$. Recall the procedure \textsc{CheckMotif} that finds $M$-matches involving $P$. It creates a list for every edge of $M$ that is not on the anchor. Each such list is of the form $E_{x,y}[s,s']$ for some vertices $x,y$ on the path $P$, where $|s -s'| \leq \delta$. Hence, the length is at most $\sigma_\delta$. The count of triples output by the call to \textsc{ListCount} is at most $\sigma^{r}_\delta$. 

So the maximum value of any $C_P$ (and hence any $X_i$) is at most $\sigma^{r}_\delta$. Let us denote this upper bound as $B$. The random variables $X_i/B$ are in $[0,1]$, and we can apply \Thm{chernoff} to $Y \eqdef \sum_{i \leq k} X_i/B$. So, $\prob[|Y - \EX[Y]| > \eps \EX[Y]] \leq 2 \exp(-\eps^2 \EX[Y]/3)$.

The events $|Y - \EX[Y]| > \eps \EX[Y]$ and $|\widehat{C} - C| > \eps C$ are identical. Observe that $\EX[Y] = \EX[\sum_{i \leq k}X_i]/B = (k/B) \cdot C/W_\delta$ (by the calculations in \Clm{est}). We choose $k \geq 3\sigma^{r}_\delta W_\delta \ln(\gamma/2)/(C\eps^2)$. Thus, $\prob[|X - \EX[X]| > \eps \EX[X]]$ is at most
\begin{align*}
& 2 \exp(-\eps^2 \EX[Y]/3)\\
& \leq 2 \exp(-(\eps^2/3) \cdot 3\sigma^{r}_\delta W_\delta \ln(\gamma/2)/(C\eps^2) \cdot (1/B) \cdot (C/W_\delta)) \\
& = \gamma.
\end{align*}
\end{proof}

We can interpret the bound above roughly as: if \textsc{Estimate} chooses at least $B W_\delta/C$ samples, it is guaranteed to give an accurate estimate. Here, $B$ refers to the maximum number of $M$-matches that share an anchor path $S$. Pessimistically, we upper bound $B$ as $\sigma^{r}_\delta$. The worst-case upper bound might be quite poor, but as we discuss in \Sec{practical}, we argue that $B W_\delta/C$ is not large in practice.

Our final theorem bounds the running time of \textsc{Estimate}.
\begin{theorem} \label{thm:est-runtime-space} 
The running time of \textsc{Estimate} is $O(m\log m + k\log m + k\sigma_\delta(|E(M)| - |V(M)|)$, where $k$ is the number of samples.
The space complexity of \textsc{Estimate} is $O(m|V(M)| + \sigma_\delta(|E(M)| - |V(M)|))$.
\end{theorem}

\begin{proof}
According to \Clm{runtime-sample}, the \textsc{Preprocess} procedure takes $O(m\log m)$ time. And running \textsc{Sample} procedure $k$ times takes $O(k \log m)$. For every sample, \textsc{CheckMotif} calls \textsc{ListCount} subroutine. As proved in \Clm{listcount-runtime}, the runtime of this procedure is $O(\sum_{i} |L_i|)$. The number of lists $L_i$ is the number of edges which are not on the specified anchor path in $M$, which is $|E(M)| - (|V(M)|-1)$. Further, each $|L_i|$ is bounded by the maximum edge multiplicity within $\delta$ time window ($\sigma_\delta$).
Therefore, the total runtime is $O(m\log m + k\log m + k\sigma_\delta(|E(M)| - |V(M)|))$

The \textsc{Preprocess} procedure stores the sampling weights for each edge in $G$ that are mapped to the selected wedge or 3-path $S$. The storage for sampling weight is $O(m|V(M)-1|)$. The \textsc{Sample} procedure needs $O(\sigma_\delta)$ for in-edges and out-edges lists.
\textsc{CheckMotif} procedure needs to keep the candidate lists of edges. The space needed is $\sum_i |L_i|$ where there are $|E(M)| - (|V(M)|-1)$ lists. And the auxiliary counter in \textsc{ListCount} subroutine needs space $max_i |L_i|$. 
So the total space complexity is $O(m|V(M)| + \sigma_\delta(|E(M)| - |V(M)|))$.
\end{proof}


\section{Brief Description of Wedge Sampling}
We begin by fixing one of the wedges in our motif as an anchor, specifically a $\langle +,> \rangle$ $\delta-$centered wedges. Similar to 3-path sampling, the wedge sampling algorithm begins with a preprocessing phase in which each edge in the input graph is assigned a non-negative weight. This weight, denoted as $w_{e,\delta}$, are calculated by $d^{+}_{u}[t,t+\delta]$ for each edge. And these weights are used to set-up a distribution for edge sampling.
During the sampling phase, we select an edge $e = (u,v,t)$ from $G$ with probability proportional to $w_{e,\delta}$. Then we uniformly sample a random edge from $\Lambda_{u}^{+}[t,t + \delta]$ that extends the sampled edge $e$ into a $\langle +,> \rangle$ $\delta-$centered wedge wedge. This method ensures that we sample uniform random wedges.
If the sampled wedge matches the anchor wedge, the \textsc{listCount} subroutine is then employed to calculate the number of instances of the motif induced by the sampled wedge.
\section{Practical considerations} \label{sec:practical}


\begin{table}[tb]
    \centering
    \scriptsize
    \caption{The maximum, average, and standard deviation of $B$ for different temporal motifs when $\delta$=4 weeks. The $B$ value is the number of motifs that share the same anchor path.}
    \label{tab:B}
    \begin{tabular}{c|c|c|c|c}
        \hline 
        \textbf{Dataset~\cite{snapnets}} & \textbf{motif} & \textbf{$B_{max}$} & \textbf{$B_{avg}$} & \textbf{$B_{std}$} \\
        \hline \hline
        \multirow{2}*{WT} & $M_\text{4-0}$ & 4.20E2 & 3.59E0 & 5.86E0 \\
        & $M_\text{4-5}$ & 8.36E4 & 1.59E2 & 6.76E+2\\
        \hline
        \multirow{2}{*}{SO} & $M_\text{4-0}$ & 9.80E1 & 1.88E0 & 1.82E0 \\
        & $M_\text{4-5}$ &  2.28E3 & 1.94E1 & 5.73E1 \\
    \hline
    \end{tabular}
\end{table}


In this section, we give numerous arguments as to why \textsc{Estimate} works in practice.
A summary of the algorithm and mathematical analysis of the previous section is the following. We can get accurate estimates to a temporal motif count $C$ with (roughly) $k = \sigma^{r}_\delta W_\delta/C$ samples. 
These are pessimistic bounds, since $\sigma_\delta$ is a (large) upper bound used in various proofs. In practice, $\sigma_\delta$ is in the thousands. 

Let us apply better ``heuristic" bounds on the sample size $k$ and the runtime. Going through the proof of \Thm{est}, we set $k = \widetilde{B} W_\delta/C$, where $\widetilde{B}$ is a reasonable upper bound on the number of motifs that share the same anchor 3-path. We do not need the exact upper bound, since (for estimation) we can afford to ignore some matches. We would like to capture as many motifs as possible, with a small value of $\widetilde{B}$. For the running time, we see that \Thm{est-runtime-space} gives a sharper bound that depends on the edge multiplicities of the edges in the match (which is trivially upper bounded by the largest multiplicity). Instead of upper bounding by $\sigma_\delta$, we can upper bound by $B_{avg}$, which is the average number of motif matches for a wedge or 3-path.

In all, our heuristic bound for the samples required is $\widetilde{B} W_\delta/C$ and the runtime (ignoring near-linear preprocess) is $\widetilde{B}\cdot B_{avg} W_\delta/C$. Our algorithm is practical due to two main reasons. 
First, the $\delta$-centered wedges or 3-paths can be sampled efficiently, because their total count $W_\delta$ is small. Second, both $\widetilde{B}$ and $B_{avg}$ are quite small in practice.
For simplicity, we only discuss $\delta$-centered 3-path here.

\noindent{\bf On why $\delta$-centered 3-paths work.}
A naive strategy for sampling 3-paths is to \emph{disregard all temporal constraints} on the 3-path during preprocessing. This would default to the static 3-path sampling of~\cite{jha2015path}. The preprocessing would be fast and sampling would be much quicker. The final \textsc{MotifCount} procedure checks for temporal constraints.
But discarding all temporal constraints during preprocessing renders the majority of the sampled 3-paths non-compliant with the specified temporal requirements, leading to only 0.01\% hit rate of \textit{valid} 3-paths. 
The static 3-path sampler does not work, because the number of samples required is infeasible.

An alternate strategy is to \emph{satisfy all temporal constraints} ($t(e_0) < t(e_1) < t(e_2) < t(e_0) + \delta$) during preprocessing. This approach eliminates unnecessary samples due to violating timestamp constraints. But preprocessing and sampling become inefficient.
Now there are dependencies among the edges $e_0$ and $e_2$, so sampling takes more time. On sampling $e_1$, we have to first sample $e_0$, and then \emph{depending on $e_0$}, set up sampling for $e_2$. 
preprocessing time complexity is now $O(m \sigma_{\delta, avg}log\sigma_{\delta, avg})$, where $\sigma_{\delta, avg}$ is the average multiplicity of an edge within the $\delta$-time window. 

The $\delta$-centered 3-paths offer a happy medium, wherein sampling is efficient and the total number of paths ($W_\delta$) remains low. 
We see that $W_\delta$ is quite comparable to the actual count $C$. Recall that the sample bound is linear in $W_\delta/C$, which in practice is quite small.

\noindent{\bf The typically low values of $\widetilde{B}$ and $B_{avg}$.} 
In \Tab{B}, we see that $B_{avg}$ is typically small and much smaller than the maximum value. This means that the practical runtime bounds are also much smaller than the pessimistic worst-case bound.
So the value of $\widetilde{B}$ is quite small, suggesting that the number of samples required in \Thm{est} is not large.

\section{Experiments}

\begin{table}[t]
\scriptsize
\caption{Temporal graph datasets used in the evaluation.}
\vspace{-1mm}
\label{tab:dataset}
\centering
\begin{tabular}[htbp]
{P{2cm}|P{0.6cm}|P{1.3cm}|P{1cm}|P{1.3cm}}
\hline
\textbf{Dataset} & \textbf{$V$} & \textbf{$|E_{temporal}|$} & \textbf{$|E_{static}|$} & \textbf{Time span (year)} \\ \hline \hline
wiki-talk (WT) & 1.1M & 7.8M & 3.3M & 6.36\\ \hline
stackoverflow (SO) & 2.6M & 63.5M & 36.2M & 7.60 \\ \hline
bitcoin (BI) & 48.1M & 113.1M & 86.8M & 7.08 \\ \hline
reddit-reply (RE) & 8.4M & 636.3M & 517.2M & 10.06 \\\hline
\end{tabular}
\end{table}

In this section, we present the performance of \THISWORK\ using both accuracy and wall clock runtime metrics. 
We evaluate the results on large-scale datasets and large $\delta-$temporal windows that have up to trillions of matches. 
Notably, the exact count algorithm fails to finish the motif counting within a week for many scenarios. 
To our knowledge, \THISWORK\ is the first sampling-based temporal motif count algorithm that targets this scale and complexity.

\subsection{Experiment Setup}
\label{sec:setup}

\noindent\textbf{Benchmark.}
We run the experiments on a collection of medium to large temporal datasets, including wiki-talk (WT), stackoverflow (SO) from SNAP~\cite{snapnets}, bitcoin (BI)~\cite{Kondor-2014-bitcoin}, and reddit-reply (RE)~\cite{hessel2016science}, listed in  Table~\ref{tab:dataset}. 
The link to those public datasets is listed below:
\begin{itemize}
    \item wiki-talk (WT): \url{https://snap.stanford.edu/data/wiki-talk-temporal.html}
    \item stackoverflow (SO) \url{https://snap.stanford.edu/data/sx-stackoverflow.html}
    \item bitcoin (BI) \url{https://www.cs.cornell.edu/~arb/data/temporal-bitcoin/}
    \item reddit-reply (RE) \url{https://www.cs.cornell.edu/~arb/data/temporal-reddit-reply/}
\end{itemize}
We evaluate the temporal motif counts for WT and SO with $\delta$ from 4W to 16W, and for BI and RE with $\delta$ = 1D. In this context, "W" represents week and "D" represents day.

\noindent\textbf{Exact Count Baseline.}
For CPU-based exact count baseline, we selected the backtracking algorithm (\textit{BT}) proposed by Machkey~\textit{et al.}~\cite{mackey2018chronological}, which does a backtracking search on chronologically-sorted temporal edges.
The original C++ implementation is single-threaded and runs for more than a week in many cases. 
We implemented a multi-threaded version of BT with OpenMP using dynamically scheduled work-stealing threads without using costly synchronization/atomic primitives.

For the GPU-based exact count baseline, we employ the state-of-the-art \textit{Everest}~\cite{yuan2023everest}, which uses the backtracking algorithm with system-level optimizations.

\noindent\textbf{Approximate Baselines.}
We compare \THISWORK\ with the state-of-the-art approximate algorithm PRESTO~\cite{sarpe2021presto}, with two variants, \textit{PRESTO-A} and \textit{PRESTO-E}. 
PRESTO~\cite{sarpe2021presto} is a sampling algorithm that runs an exact motif count algorithm on sampled intervals to get estimated results. 
is similar to IS~\cite{liu2019sampling}, but does not require partitioning all edges into non-overlapping windows.  Instead, it leverages uniform sampling. 
It provides two variants, \textit{PRESTO-A} and \textit{PRESTO-E}. 
We use their open-source implementation from \cite{PRESTO_impl}.
We run \textit{PRESTO-A} and \textit{PRESTO-E} with a scaling factor of sampling window size $c=1.25$. The number of samples configured such that its runtime is around 10$\times$ slower than \THISWORK.
We omit the comparison to IS~\cite{liu2019sampling} and ES~\cite{wang2020efficient} because PRESTO outperforms them and ES code is limited to motifs with 4 or fewer edges.

\noindent\textbf{Metric}
The accuracy of approximate algorithms is defined as $\frac{|C-\hat{C}|}{C}$, where $C$ is the exact count from Everest, and $\hat{C}$ is the estimated count. And error = 1- accuracy. We run 5 times to get the average (avg) and standard deviation (std) of errors.

\noindent\textbf{Hardware Platform.}
We run Everest~\cite{yuan2023everest} on a single server-grade NVIDIA A40 GPU with 10k CUDA cores.
For the remaining baselines and \THISWORK\ written in C++, we run the experiments on an AMD EPYC 7742 64-Core CPU with 64MB L2 cache, 256MB L3 cache, and 1.5TB DRAM memory. We implement multi-threading using the OpenMP library. All CPU-based algorithms run with 32 threads. 

\subsection{Results}
In \Tab{count}, we list the number of temporal motifs for $M_i$ in \Fig{temp_motifs} with various datasets and $\delta$. We mark it as "timeout" if Everest cannot finish in 1 day. We are interested in cases where instances reach a trillion scale.

\begin{table*}[htb]
    \caption{Exact motif counts for $M_i$. \textit{timeout} means Everest (GPU) could not finish counting in 1 day.}
    \vspace{-2mm}
    \label{tab:count}
    \centering
    \scriptsize
    \begin{tabular}{c|c|c|c|c|c|c|c|c|c}
        \hline
Dataset & $\delta$ & $M_\text{3-0}$ & $M_\text{3-1}$ & $M_\text{4-0}$ &$M_\text{4-1}$ & $M_\text{4-2}$ & $M_\text{4-3}$ & $M_\text{4-4}$ & $M_\text{4-5}$ \\
\hline \hline
\multirow{2}{*}{WT} & 4W & 6.4E11 & 8.9E11 & 3.0E8 & 8.8E12 & 2.8E10 & 1.3E11 & 4.3E11 & 1.2E9 \\ \cline{2-10}
& 8W & 9.4E13 & 1.3E13 & 1.5E9 & timeout & 3.4E12 & 4.5E12 & 8.4E13 & 2.3E10\\ \hline
SO & 16W & 3.3E11 & 3.7E11 & 4.0E9 & 7.9E11 & 5.5E10 & 3.6E11 & 7.0E11 & 5.7E9\\ \hline
BI & 1D & 1.1E14 & 3.3E13 & 6.5E9 & timeout & 6.7E12 & 5.2E13 & 2.3E13 & 6.1E10\\ \hline
RE & 1D & 5.4E12 & 1.3E13 & 2.8E9 & 1.5E13 & 2.4E11 & 1.6E12 & 1.0E13 & 1.0E11 \\ \hline
    \end{tabular}
\end{table*}

\begin{table}[htbp]
\centering
\scriptsize
\caption{Runtime(s) for Everest (GPU) and \THISWORK(CPU), and error of \THISWORK. The speedup value is the runtime of \THISWORK\ (CPU) over Everest (GPU). We mark it as \textit{timeout} if it cannot finish within one day. We mark the speedup and error as \textit{No Exact} for the case that Everest timeout. The error(\%) is represented as avg $\pm$ std.}
\label{tab:runtime-BT-this}
\begin{tabular}{C{0.4cm}|c|C{0.6cm}|C{1cm}|C{1.1cm}|C{1cm}|C{1.2cm}}
 \textbf{$G$} & \textbf{$\delta$} & \textbf{Motif} & \textbf{Everest} & \textbf{\THISWORK} & \textbf{Speedup ($\times$)} & \textbf{Error(\%)} \\ \hline \hline
\multirow{16}{*}{WI} & \multirow{8}{*}{4W} & $M_\text{3-0}$ & 221.0 & 10.9 & 20.2 & 1.3$\pm$0.4 \\ \cline{3-7} 
 &  & $M_\text{3-1}$ & 309.9 & 10.9 & 28.3 & 1.9$\pm$1.4 \\ \cline{3-7} 
 &  & $M_\text{4-0}$ & 7.8 & 4.8 & 1.6 & 0.7$\pm$0.5 \\ \cline{3-7} 
 &  & $M_\text{4-1}$ & 6989.0 & 12.8 & 545.6 & 7.2$\pm$5.9 \\ \cline{3-7} 
 &  & $M_\text{4-2}$ & 22.3 & 14.4 & 1.6 & 3$\pm$1.8 \\ \cline{3-7} 
 &  & $M_\text{4-3}$ & 80.2 & 13.2 & 6.1 & 0.9$\pm$0.4 \\ \cline{3-7} 
 &  & $M_\text{4-4}$ & 186.5 & 12.9 & 14.5 & 7.3$\pm$5.5 \\ \cline{3-7} 
 &  & $M_\text{4-5}$ & 14.1 & 13.6 & 1.0 & 1.8$\pm$1.1 \\ \cline{2-7} 
 & \multirow{8}{*}{8W} & $M_\text{3-0}$ & 22673.0 & 10.8 & 2093.5 & 1.1$\pm$0.4 \\ \cline{3-7} 
 &  & $M_\text{3-1}$ & timeout & 10.6 & No Exact &  No Exact \\ \cline{3-7} 
 &  & $M_\text{4-0}$ & 44.7 & 4.7 & 9.6 & 0.6$\pm$0.3 \\ \cline{3-7} 
 &  & $M_\text{4-1}$ & timeout & 12.9 &  No Exact &  No Exact \\ \cline{3-7} 
 &  & $M_\text{4-2}$ & 1022.1 & 14.1 & 72.3 & 2.1$\pm$1.8 \\ \cline{3-7} 
 &  & $M_\text{4-3}$ & 1661.8 & 13.6 & 122.6 & 0.5$\pm$0.4 \\ \cline{3-7} 
 &  & $M_\text{4-4}$ & 25605.4 & 13.0 & 1975.7 & 3.5$\pm$3.1 \\ \cline{3-7} 
 &  & $M_\text{4-5}$ & 98.9 & 13.9 & 7.1 & 2.5$\pm$1.7 \\ \hline
\multirow{8}{*}{SO} & \multirow{8}{*}{16W} & $M_\text{3-0}$ & 697.3 & 30.8 & 22.7 & 5.6$\pm$2.9 \\ \cline{3-7} 
 &  & $M_\text{3-1}$ & 668.3 & 31.3 & 21.3 & 12.8$\pm$6.5 \\ \cline{3-7} 
 &  & $M_\text{4-0}$ & 1873.8 & 36.5 & 51.3 & 0.6$\pm$0.1 \\ \cline{3-7} 
 &  & $M_\text{4-1}$ & 13377.0 & 38.7 & 346.0 & 4.4$\pm$5.3 \\ \cline{3-7} 
 &  & $M_\text{4-2}$ & 1802.7 & 37.9 & 47.5 & 13.3$\pm$7.3 \\ \cline{3-7} 
 &  & $M_\text{4-3}$ & 4669.9 & 34.6 & 134.9 & 2.9$\pm$2 \\ \cline{3-7} 
 &  & $M_\text{4-4}$ & 8069.0 & 195.1 & 41.4 & 15.4$\pm$7.5 \\ \cline{3-7} 
 &  & $M_\text{4-5}$ & 6385.9 & 36.6 & 174.5 & 2$\pm$2.1 \\ \hline
\multirow{8}{*}{BI} & \multirow{8}{*}{1D} & $M_\text{3-0}$ & 37665.9 & 43.7 & 861.3 & 2.8$\pm$1.2 \\ \cline{3-7} 
 &  & $M_\text{3-1}$ & 9598.9 & 43.5 & 220.9 & 2.9$\pm$1.5 \\ \cline{3-7} 
 &  & $M_\text{4-0}$ & 240.6 & 50.8 & 4.7 & 0.4$\pm$0.3 \\ \cline{3-7} 
 &  & $M_\text{4-1}$ & timeout & 50.6 &  No Exact &  No Exact \\ \cline{3-7} 
 &  & $M_\text{4-2}$ & 5190.2 & 49.5 & 104.9 & 2.7$\pm$2.1 \\ \cline{3-7} 
 &  & $M_\text{4-3}$ & 23143.1 & 47.5 & 487.4 & 0.6$\pm$0.2 \\ \cline{3-7} 
 &  & $M_\text{4-4}$ & 22867.4 & 46.9 & 488.1 & 9.4$\pm$3.6 \\ \cline{3-7} 
 &  & $M_\text{4-5}$ & 1244.8 & 48.0 & 25.9 & 2.3$\pm$1.1 \\ \hline
\multirow{8}{*}{RE} & \multirow{8}{*}{1D} & $M_\text{3-0}$ & 2841.2 & 183.0 & 15.5 & 5.8$\pm$4.2 \\ \cline{3-7} 
 &  & $M_\text{3-1}$ & 5761.1 & 271.4 & 21.2 & 3.5$\pm$3 \\ \cline{3-7} 
 &  & $M_\text{4-0}$ & 163.3 & 213.6 & 0.8 & 0.3$\pm$0.2 \\ \cline{3-7} 
 &  & $M_\text{4-1}$ & 18967.5 & 249.0 & 76.2 & 26.1$\pm$12.7 \\ \cline{3-7} 
 &  & $M_\text{4-2}$ & 188.6 & 215.1 & 0.9 & 9.4$\pm$5 \\ \cline{3-7} 
 &  & $M_\text{4-3}$ & 764.5 & 221.6 & 3.4 & 3.9$\pm$3.1 \\ \cline{3-7} 
 &  & $M_\text{4-4}$ & 10472.2 & 353.5 & 29.6 & 11.1$\pm$10.8 \\ \cline{3-7} 
 &  & $M_\text{4-5}$ & 273.5 & 220.3 & 1.2 & 3.2$\pm$3.1 \\ \hline

\end{tabular}
\end{table}

\noindent\textbf{Comparison with Exact Algorithms.}
We present a runtime comparison between Everest (GPU)~\cite{yuan2023everest,Everest_impl} and \THISWORK\ (CPU) in \Tab{runtime-BT-this}. The detailed runtime for BT (CPU) is omitted due to space constraints; typically, BT runs about 10$\times$ slower than Everest.
As the number of instances escalates to the hundreds of trillions (as shown in \Tab{count}), Everest, even on the NVIDIA A40 GPU, struggles to scale effectively. In some scenarios, it fails to complete within a day. In contrast, \THISWORK\ consistently finishes in under 6 minutes, achieving an average speedup of 33x, and in some cases, up to 2000$\times$ over Everest, and an average speedup of 170$\times$ over BT.

The table also highlights the error rates for \THISWORK, which are typically below $10\%$ with minimal variance. This demonstrates that \THISWORK\ is both fast and accurate.

\noindent\textbf{Comparison with Approximate Algorithms.}
For the approximate algorithms, both estimation accuracy and runtime are important metrics. 
We assess accuracy and runtime against $PRESTO$~\cite{sarpe2021presto}, a state-of-the-art prior work. 
To ensure a conservative comparison, we configure the number of samples in the PRESTO implementation~\cite{PRESTO_impl} such that PRESTO's runtime is at least 10 times longer than \THISWORK.
Both run 32 threads on CPU. 
Table~\ref{tab:approx} illustrates that \THISWORK\ consistently surpasses PRESTO in terms of runtime and is nearly always more accurate.
The lower accuracy of PRESTO can be attributed to its procedure of uniformly partitioning time intervals with a length of $c\delta$, where $c$ represents the sampling window size. This approach assumes uniformity in time intervals, which may not hold true for skewed input graphs, leading to inaccuracies in estimation. 
Moreover, PRESTO relies on the BT algorithm~\cite{mackey2018chronological} as a subroutine for processing edges within sampled $c\delta$-length time windows. 
However, as previously demonstrated, the BT algorithm exhibits scalability issues with large $\delta$ values, resulting in significant runtime even for a single sample.
Additionally, workload imbalance across multiple threads can further hamper PRESTO's performance, particularly with skewed input graphs.
In contrast, \THISWORK\ mitigates the skewed graph connectivity during preprocessing by meticulously constructing sampling weights, resulting in higher accuracy. 
For runtime, \THISWORK\ leverages the \textsc{Derivecnts} algorithm, which counts subgraphs in linear time without enumeration, resulting in improved performance.

\begin{table}[t]
\centering
\scriptsize
\caption{Runtime (s) and relative error (\%) of all approximate algorithms on various datasets and temporal motifs. The lowest error is highlighted in \colorbox{Green1}{green} block. The smallest runtime is highlighted in \colorbox{Blue}{blue} block.  
We mark the error \textit{NE} (No Exact) when Everest timeout and no exact counts are available. 
If the program runs out of memory (OOM), it will be \textit{killed} by Linux. \THISWORK\ is the fastest and the most accurate in most cases.}
\label{tab:approx}
\begin{tabular}{P{0.8cm}|P{0.6cm}|P{0.8cm}|P{0.6cm}|P{0.8cm}|P{0.6cm}|P{0.8cm}|P{0.6cm}}
\hline
\multirow{2}{*}{ \textbf{Dataset} } & \multirow{2}{*}{ \textbf{Motif} } & \multicolumn{2}{c|}{\textbf{PRESTO-A}} & \multicolumn{2}{c|}{\textbf{PRESTO-E}} & \multicolumn{2}{c}{\textbf{\THISWORK}} \\ 
\cline{3-8}
& & \textbf{Time (s)} & \textbf{Error} & \textbf{Time (s)} & \textbf{Error} & \textbf{Time (s)} & \textbf{Error} \\ \hline\hline
\multirow{8}{*}{\shortstack{WT\\ $\delta$=4W}} & $M_\text{3-0}$ & 2.5E3 & 18.9\% & 9.3E3 & 36.2\% & \cellcolor{Blue}\textbf{1.2E1} & \cellcolor{Green1}\textbf{1.3\%} \\ \cline{2-8}
& $M_\text{3-1}$ & 8.4E3 & 221.0\% & 1.9E4 & 4.2\% & \cellcolor{Blue}\textbf{1.1E1} & \cellcolor{Green1}\textbf{1.9\%} \\ \cline{2-8}
& $M_\text{4-0}$ & 4.3E3 & 3.0\% & 4.3E3 & 3.1\% & \cellcolor{Blue}\textbf{4.8E0} & \cellcolor{Green1}\textbf{0.7\%} \\ \cline{2-8}
& $M_\text{4-1}$ & 3.3E4 & 14.3\% & 2.0E5 & 90.5\% & \cellcolor{Blue}\textbf{1.3E1} & \cellcolor{Green1}\textbf{7.2\%} \\ \cline{2-8}
 & $M_\text{4-2}$ & 2.5E2 & 123.4\% & 2.8E2 & \cellcolor{Green1}\textbf{3.0\%} & \cellcolor{Blue}\textbf{2.5E1} & 12.2\% \\ \cline{2-8}
 & $M_\text{4-3}$ & 4.2E2 & 17.4\% & 1.4E3 & \cellcolor{Green1}\textbf{0.6\%} & \cellcolor{Blue}\textbf{1.3E1} & 2.0\% \\ \cline{2-8}
 & $M_\text{4-4}$ & 3.7E3 & 30.3\% & 8.9E3 & 35.4\% & \cellcolor{Blue}\textbf{1.2E1} & \cellcolor{Green1}\textbf{5.5\%} \\ \cline{2-8}
 & $M_\text{4-5}$ & 1.3E2 & 62.7\% & 1.2E2 & 43.2\% & \cellcolor{Blue}\textbf{1.4E1} & \cellcolor{Green1}\textbf{1.8\%} \\ \hline
\multirow{8}{*}{\shortstack{SO\\ $\delta$=16W}} & $M_\text{3-0}$ & 1.9E3 & 9.8\% & 9.8E3 & 35.4\% & \cellcolor{Blue}\textbf{2.7E1} & \cellcolor{Green1}\textbf{2.2\%} \\ \cline{2-8}
& $M_\text{3-1}$ & 2.6E2 & 103.5\% & 1.1E4 & 19.9\% & \cellcolor{Blue}\textbf{3.1E1} & \cellcolor{Green1}\textbf{12.8\%} \\ \cline{2-8}
 & $M_\text{4-0}$ & killed & N/A & killed & N/A & \cellcolor{Blue}\textbf{3.7E1}  & \cellcolor{Green1}\textbf{0.2\%} \\ \cline{2-8}
  & $M_\text{4-1}$ & 1.3E4 & 5.0\% & 2.3E4 & 103.0\% & \cellcolor{Blue}\textbf{3.9E1}  & \cellcolor{Green1}\textbf{4.4\%} \\ \cline{2-8}
 & $M_\text{4-2}$ & killed & N/A & killed & N/A & \cellcolor{Blue}\textbf{4.0E1} & \cellcolor{Green1}\textbf{7.2\%}  \\  \cline{2-8}
 & $M_\text{4-3}$ & 5.8E3 & 5.1\% & 7.7E3 & 57.8\% & \cellcolor{Blue}\textbf{3.1E1} & \cellcolor{Green1}\textbf{1.2\%} \\ \cline{2-8}
 & $M_\text{4-4}$ & 8.8E3 & 38.7\% & 1.1E4 & 36.0\% & \cellcolor{Blue}\textbf{2.4E2} & \cellcolor{Green1}\textbf{5.8\%} \\ \cline{2-8}
 & $M_\text{4-5}$ & 8.4E3 & 13.8\% & 1.0E4 & 22.9\% & \cellcolor{Blue}\textbf{3.7E1} & \cellcolor{Green1}\textbf{2.7\%} \\ \hline
\multirow{8}{*}{\shortstack{BI\\ $\delta$=1D}} & $M_\text{3-0}$ & 9.8E2 & 91.5\% & 3.6E4 & 41.2\% & \cellcolor{Blue}\textbf{4.4E1} & \cellcolor{Green1}\textbf{2.8\%} \\ \cline{2-8}
& $M_\text{3-1}$ & 1.1E3 & 84.7\% & 2.1E3 & 85.4\% & \cellcolor{Blue}\textbf{4.4E1} & \cellcolor{Green1}\textbf{2.9\%} \\ \cline{2-8}
 & $M_\text{4-0}$ & 3.8E2 & 12.9\% & 1.0E3 & 10.7\% & \cellcolor{Blue}\textbf{5.1E1} & \cellcolor{Green1}\textbf{0.4\%} \\ \cline{2-8}
 & $M_\text{4-1}$ & 6.3E4 & NE & 4.5E4 & NE & \cellcolor{Blue}\textbf{5.1E1} & NE \\ \cline{2-8}
 & $M_\text{4-2}$ & 8.9E2 & 33.5\% & 4.0E3 & 37.0\% & \cellcolor{Blue}\textbf{4.9E1} & \cellcolor{Green1}\textbf{2.7\%} \\ \cline{2-8}
 & $M_\text{4-3}$ & 1.2E3 & 65.5\% & 8.0E3 & 47.6\% & \cellcolor{Blue}\textbf{4.7E1} & \cellcolor{Green1}\textbf{0.6\%} \\ \cline{2-8}
 & $M_\text{4-4}$ & 4.7E3 & 33.0\% & 6.3E3 & 64.5\% & \cellcolor{Blue}\textbf{4.7E1} & \cellcolor{Green1}\textbf{9.4\%} \\ \cline{2-8}
 & $M_\text{4-5}$ & 2.8E2 & 31.6\% & 7.3E2 & 9.0\% & \cellcolor{Blue}\textbf{4.8E1} & \cellcolor{Green1}\textbf{2.3\%} \\ \hline
\multirow{8}{*}{\shortstack{RE\\ $\delta=1D$}} & $M_\text{3-0}$ & 9.1E3 & 95.3\% & 1.7E3 & 85.5\% & \cellcolor{Blue}\textbf{1.8E2} & \cellcolor{Green1}\textbf{5.8\%} \\ \cline{2-8}
 & $M_\text{3-1}$ & 1.8E3 & 77.6\% & 4.8E3 & 25.2\% & \cellcolor{Blue}\textbf{2.7E2} & \cellcolor{Green1}\textbf{3.5\%} \\ \cline{2-8}
 & $M_\text{4-0}$ & 4.9E2 & 7.5\% & 2.1E3 & 4.3\% & \cellcolor{Blue}\textbf{2.1E2} & \cellcolor{Green1}\textbf{0.3\%} \\ \cline{2-8}
 & $M_\text{4-1}$ & 5.5E4 & \cellcolor{Green1}\textbf{25.1\%} & 2.6E3 & 94.8\% & \cellcolor{Blue}\textbf{2.5E2} & 26.1\% \\ \cline{2-8}
 & $M_\text{4-2}$ & 1.8E3 & 64.8\% & 1.6E3 & 75.8\% & \cellcolor{Blue}\textbf{2.2E2} & \cellcolor{Green1}\textbf{9.4\%} \\ \cline{2-8}
 & $M_\text{4-3}$ & 1.6E3 & 112.3\% & 3.4E4 & 104.7\% & \cellcolor{Blue}\textbf{2.2E2} & \cellcolor{Green1}\textbf{3.9\%} \\ \cline{2-8}
 & $M_\text{4-4}$ & 6.2E4 & 38.3\% & 1.8E3 & 96.4\% & \cellcolor{Blue}\textbf{2.2E2} & \cellcolor{Green1}\textbf{11.1\%} \\ \cline{2-8}
 & $M_\text{4-5}$ & 1.2E3 & 10.2\% & 4.5E3 & 18.73\% & \cellcolor{Blue}\textbf{2.2E2} & \cellcolor{Green1}\textbf{3.2\%} \\ \hline
\end{tabular}
\end{table}

\begin{figure}[t]
    \centering
    \begin{subfigure}{0.48\linewidth}
       \includegraphics[width=\linewidth]{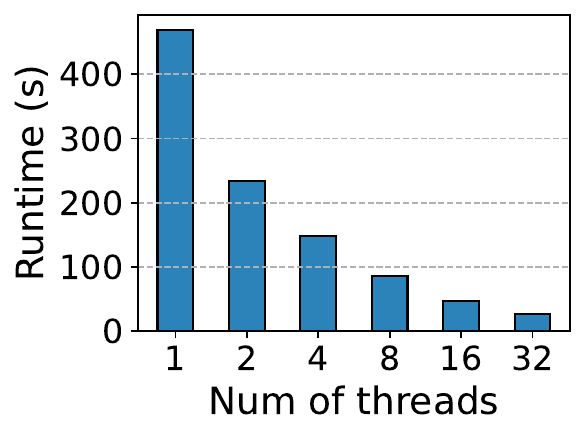}  \vspace{-5mm}
       \caption{BI}
    \end{subfigure}
    \begin{subfigure}{0.48\linewidth}
       \includegraphics[width=\linewidth]{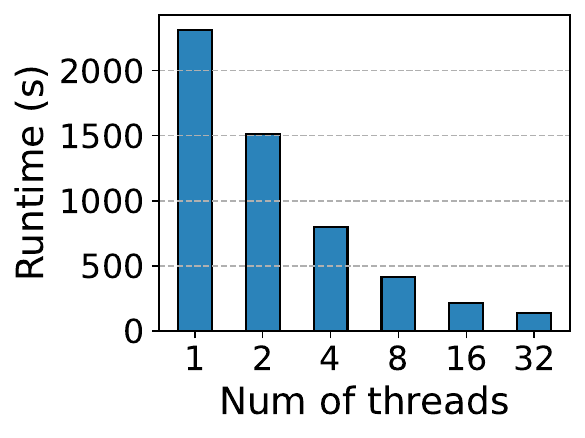}  \vspace{-5mm}
       \caption{RE}
    \end{subfigure}
    \vspace{-2mm}
    \caption{Runtime (in seconds) of the multi-threaded \THISWORK\ algorithm by varying the number of threads.}
    \label{fig:multi-thread}
\end{figure}

\noindent\textbf{Runtime Scalability.}
According to Algorithm~\ref{algo:overall}, the \textit{preprocess} step is conducted on each temporal edge, while the \textsc{sample} and \textsc{checkMotif} function are inside a \textit{for} loop with $k$ iterations. 
Importantly, all the steps in \THISWORK\ are amenable to parallelism.
Leveraging this characteristic, we implement a multi-threaded path-sampling algorithm using work-stealing OpenMP threads.  
The runtime scalability of \THISWORK\ is shown in Figure~\ref{fig:multi-thread}. 
We observe a near-linear runtime reduction for multi-threads. 
With 32-thread path-sampling algorithm, the speedup over 32-thread BT exact count algorithm is as large as four orders of magnitude (Table~\ref{tab:runtime-BT-this}).
This parallel implementation significantly improves the efficiency of \THISWORK, allowing it to harness the power of modern multi-core processor micro-architecture, and efficiently process large-scale temporal graphs in a scalable fashion.

\noindent\textbf{Insights from Motif Counts.} 
The count of multi-graph motif instances is significantly higher than that of non-multi-graph motif instances. For example, in the WT dataset with $\delta=4W$, the count for the multi-graph 4-cycle motif, $M_\text{4-2}$, is 400 billion, while the count for the simple directed 4-cycle $M_\text{4-0}$ is only 300 million. This means the multi-graph 4-cycle motif count is a thousand times larger than the non-multi-graph 4-cycle count, highlighting the dramatic increase in complexity when dealing with multi-graph temporal motifs and underscoring the limitations of previous methods.

When analyzing multi-graph triangle motifs $M_{\text{3-0}}$ and $M_{\text{3-1}}$, changing the order and direction of edges results in different motif counts. For instance, in the WT dataset with $\delta=4W$, the count for $M_{\text{3-1}}$ is approximately 30\% higher than that of $M_{\text{3-0}}$. Conversely, in the BI dataset with $\delta=1D$, the count for $M_{\text{3-1}}$ is an order of magnitude lower than that of $M_{\text{3-0}}$.
The fast and accurate nature of our algorithm enables getting these interesting insights rapidly to data scientists that can be further used in various downstream tasks (\textit{e.g.,} explaining graph neural networks~\cite{chen2024tempme}).

\section{Conclusion}
This paper presented \THISWORK, a path sampling algorithm to estimate temporal motif counts accurately and efficiently to address the scalability challenge. 
\THISWORK\ exhibits a speedup of up to three orders of magnitude over the exact count algorithm with a relative error less than 10\%, outperforming both exact and approximate state-of-the-art techniques. 

\section*{Acknowledgement}
This work is partially supported by NSF CCF-1740850, CCF-1839317, CCF-2402572, and DMS-2023495.

\bibliographystyle{IEEEtran}
\bibliography{refs}


\end{document}